\documentclass[%
 preprint,
 amsmath,amssymb,
 aps,
]{revtex4-1}
\usepackage{amsmath}
\usepackage{color}
\usepackage{tabularx}
\usepackage{multirow}
\usepackage{graphicx}
\usepackage{hyperref}
\usepackage{dcolumn}
\usepackage{bm}

\begin{document}

\title{One dimensional models for supercritical and subcritical transitions in rotating convection}

\author{Sutapa Mandal}
\affiliation{
Department of Mathematics, National Institute of Technology, Durgapur 713209, India
}
\author{Snehashish Sarkar}
\affiliation{
Department of Mathematics, National Institute of Technology, Durgapur 713209, India
}%
\author{Pinaki Pal}
\affiliation{
Department of Mathematics, National Institute of Technology, Durgapur 713209, India
}%

\date{\today}

\begin{abstract}
Numerous study on natural and man made systems including rotating convection report the phenomena of supercritical and subcritical transitions from one state to another with the variation of relevant control parameters. However, the complexity of the rotating convection system even under the idealized  Rayleigh-B\'enard geometry, hindered the simplest possible description of these transitions to convection. Here we present an one dimensional description of the stationary subcritical and supercritical transitions to rotating Rayleigh-B\'enard convection both for rigid and free-slip boundary conditions. The analysis of the one dimensional models and performance of three dimensional direct numerical simulations of the system show qualitatively similar results in a wide region of the parameter space. A brief discussion on time dependent convection of overstable origin is also presented.
\end{abstract}

\maketitle


\section{Introduction}

The phenomenon of transition from one state to the other in a system with the variation of control parameters is ubiquitous~\cite{glatzmaier:1999,knobloch:JFM_1981,davidson:ARFM_1999}, and is characterized by an order parameter. Examples include transition from conduction to convection~\cite{chandra:book,mkv:book}, vapor to liquid~\cite{elliott_PRL:2002} and laminar to turbulent~\cite{loshe:RMP2009} in fluids,  graphite to diamond~\cite{xie:2014}, ferromagnetic to paramagnetic~\cite{eich:2017} etc. Depending on the nature of the system, the order parameter may exhibit continuous or discontinuous transitions at a critical value of the parameter. These transitions are  connected with the supercritical and subcritical bifurcations of the system respectively, and are named accordingly. 

The current paper deals with the problem of thermal convection in the presence of rotation. The phenomenon of thermal convection is observed in wide variety of natural as well as man made systems including geophysical~\cite{glatzmaier:1999}, astrophysical~\cite{davidson:ARFM_1999}, oceanic~\cite{marshall:1999},  liquid metals~\cite{kirillov:1995,Olson:2001,kaplan:2017}, liquid crystal~\cite{hurle:1972,halperin:1974} etc., and it is one of the key factors governing the dynamics there. The richness of dynamics in such systems in the accessible parameter ranges attracted the attentions of the researchers for long, and kept the field an active area of research~\cite{bodenschatz:ARFM_2000,net:2008,verma:2017,nandukumar:2019,cooper:2020,ghosh_POF:2020,mandal:POF_2022,mandal_POF:2023}.  To unfold the complexity of thermal convection, researchers often rely of the plane layer Rayleigh B\'enard convection (RBC) model for the investigation of convective phenomena like instabilities~\cite{busse:JFM_1971,clever:POF_1990,bodenschatz:ARFM_2000,hirdesh:POF_2013,dan:EPJB_2014}, patterns~\cite{cross:RMP_1993,rebecca:book,pal:PRE_2013}, chaos~\cite{knobloch_JFM:1986,paul:2011,nandu:EPL_2015}, heat transfer~\cite{rumford:1797,thual:1992,Olson:2001,ahlers:RMP_2009}, turbulence~\cite{mkv:book,manneville_book:2010} etc. For a fixed investigation domain, RBC is completely described by two parameters, namely, the Rayleigh number ($\mathrm{Ra}$, measures the vigor of the buoyancy force) and the Prandtl number ($\mathrm{Pr}$).

The study of RBC consisting of a horizontal layer of unstably stratified fluid kept between two horizontal plates, maintained at constant temperatures by heating from below and cooled from above~\cite{chandra:book,bodenschatz:ARFM_2000,loshe:RMP2009,Ecke:ARFM2023} for more than a century has not only improved the understanding of thermal convection in wide variety of fluids but also contributed significantly in developing the subjects like hydrodynamic and hydromagnetic instabilities~\cite{chandra:book}, pattern formation~\cite{rebecca:book,pal:PRE_2013}, and nonlinear dynamics~\cite{stogatz:book}. However, for geophysical~\cite{evonuk:PSS_2007}, oceanic~\cite{marshall:1999} and astrophysical~\cite{barker:AJ2014,Miesch:2000} systems where convection occurs in the presence of rotation, rotating  Rayleigh B\'enard convection (RRBC) provide a better model. The presence of rotation introduces centrifugal as well as Coriolis forces into the system along with the buoyancy force and makes the problem more complex compared to its non-rotating counterpart~\cite{chandra:book,rossby:JFM_1969,Ecke:ARFM2023}.

In this paper, we consider RRBC of low Prandtl number fluids, description of which needs one more dimensionless parameter, namely, the Taylor number ($\mathrm{Ta}$, measures the strength of the rotation about a vertical axis) along with $\mathrm{Ra}$ and $\mathrm{Pr}$. Chandrasekhar~\cite{chandra:book} developed the linear theory based on normal mode analysis to determine the critical Rayliegh number ($\mathrm{Ra}_c$) and wave number ($\mathrm{k}_c$) for the onset of RRBC both in the presence of free-slip and no-slip velocity boundary conditions. Linear theory results show that the rotational constraint inhibits convection by pushing the critical Rayleigh number for the onset of stationary as well as oscillatory convection towards higher $\mathrm{Ra}$ which is also experimentally supported~\cite{nakagawa:1955,Fultz:1955}. Note that for higher $\mathrm{Ta}$ and $\mathrm{Pr} < 0.677$, the effect of rotational constraint is reduced and time dependent overstable solutions are observed at the onset. Beyond the onset of convection, RRBC have been extensively investigated theoretically~\cite{veronis:1959,Eltayeb:1972,clune:1993,julien_POF:1999},numerically~\cite{Julien_JFM:1996,Kunnen_PRE:2006,Julien_PRL:2010,Maity:EPL2013,Maity:POF2014} and experimentally~\cite{rossby:JFM_1969,Sakai_JFM:1997,Vorobieff:JFM2002,bajaj:PRE_2002,Vasil_PNAS:2021}. These studies revealed several interesting properties of rotating convection related to instabilities,  bifurcations,  pattern dynamics,  and  turbulence.  

However, here we focus on the primary instability and the related flow patterns near the onset of convection. Of particular interest is the subcritical convection leading to finite amplitude solution at the onset. The existence of such subcritical convection was first theoretically shown in RRBC with free slip boundary conditions using perturbation method~\cite{veronis:1959} and low dimensional modeling~\cite{veronis:1966}. 
On the other hand, first experimental observation of subcritical rotating convection was reported by Rossby~\cite{rossby:JFM_1969}. Subsequently, Clever and Busse~\cite{clever1:JFM_1979} numerically examined subcritical convection in low Prandtl number fluids in the presence of rigid boundaries. The theoretical analysis of Clune and Knobloch~\cite{clune:1993} based on weakly nonlinear theory, followed by the simultaneous experimental and numerical study of Bajaj et al.~\cite{bajaj:PRE_2002} also provided great insight on the phenomenon of subcritical convection and associated finite amplitude solution at the onset of RRBC.

Recently, in an extensive three dimensional direct numerical simulations with rigid boundaries, Mandal et al.~\cite{mandal:POF_2022} identified the region of the parameter space for the observation of finite amplitude solution at the onset of convection. Along with the direct numerical simulations, a low dimensional model ($22$-dimensional) also used to analyze the finite amplitude solutions at the onset and the origin of it is connected to the subcritical pitchfork bifurcation of the basic conduction state.  The investigation also revealed that the rotation promotes subcriticality, while, Prandtl number inhibits it in the stationary convection regime. However, a simplified description of the phenomenon is still missing, due to the inherent complexity of the RRBC system. 

Here we revisit the problem of subcritical rotating convection in low Prandtl number fluids with the objective of providing a simplest possible description of the phenomena. The study is performed using RRBC model with both rigid and free-slip boundary conditions by varying the Prandtl number in the range $0 < \mathrm{Pr} \leq 0.6$. For rigid boundaries, the Taylor number is varied in the range $0<\mathrm{Ta}\leq 5 \times 10^4$ and for free-slip boundaries it is varied in the range $0<\mathrm{Ta}\leq 10^4$.  Two one dimensional models, one each for rigid and free-slip boundary conditions are derived using Galerkin projection and adiabatic elimination process for that purpose, which nicely captures the phenomena of subcritical convection and related transitions. We also perform direct numerical simulations (DNS) of the system in the said parameter regime. The model and DNS results show qualitative match. Additionally, we also investigate the onset of overstable convection in this paper.

\section{Problem Formulation}
\subsection{Physical system and governing equations}
Standard plane layer Rayleigh-B\'enard convection system  consisting of a thin horizontal layer of Newtonian fluid of thickness $d$, kinematic viscosity $\nu$, thermal diffusivity $\kappa$ and coefficient of volume expansion $\alpha$ confined between two perfectly thermally conducting horizontal plates is considered for the study. The convective motion is driven by the buoyancy force generated due to the thermal gradient between the upper and lower plates, maintained at constant temperatures $T_u$ and $T_l$ respectively, with $\Delta T = T_l - T_u > 0$. The system is rotated about the vertical axis with an angular velocity $\boldsymbol{\Omega}$ ($= \Omega {\bf {\hat{e}}}_z$, ${\bf {\hat{e}}}_z$ being the vertically upward unit vector). As $\Delta T$ crosses a critical value, the convective motion of the fluid in the presence of rotation is described by the following set of dimensionless Boussinesq-Oberbeck~\cite{Oberbeck:1879,Boussinesq:1903} equations with respect to a frame of reference co-rotating with the system:

\begin{subequations}
\begin{gather}
\frac{\mathcal{D}{\bf{u}}}{\mathcal{D}t} = -\boldsymbol{\nabla}{\pi} + \nabla^2{\bf{u}} + \mathrm{Ra} \theta {\bf{\hat{e}}}_z
+\sqrt{\mathrm{Ta}}({\bf u} \times {\bf{\hat{e}}}_z), \label{eq:momentum} \\
\frac{\mathcal{D}{\theta}}{\mathcal{D}t} = \frac{1}{\mathrm{Pr}}\left[u_z+\nabla^2\theta\right] \label{eq:heat},\\
\boldsymbol{\nabla \cdot} {\bf{u}} = 0, \label{eq:div_free}
\end{gather}
\end{subequations}
where $\frac{\mathcal{D}}{\mathcal{D}t}\equiv \frac{\partial}{\partial t} + ({\bf{u}}\boldsymbol{\cdot}\boldsymbol{\nabla})$ represents the material derivative, and ${\bf u}(x,y,z,t) = (u_x, u_y, u_z)$, $\theta(x,y,z,t)$ and $\pi(x,y,z,t)$ are the convective velocity,  temperature and pressure fields respectively. Note that the convective pressure field $\pi(x,y,z,t)$ includes the contribution of the centrifugal acceleration. The scales $d$, $d^2/\nu$, and $\Delta T \nu /\kappa$ for length, time and temperature, respectively, are used to make the equations (\ref{eq:momentum}) - (\ref{eq:div_free}) dimensionless. The Rayleigh and Prandtl numbers are defined by $\mathrm{Ra} = \alpha g \Delta T d^3/(\nu\kappa)$ and $\mathrm{Pr} = \nu/\kappa$, where $g$ is the acceleration due to gravity. Another parameter called the reduced Rayleigh number ($\mathrm{r}$) is used subsequently and it is defined by $\mathrm{r} = \mathrm{Ra}/\mathrm{Ra}_c$, where $\mathrm{Ra}_c$ is the critical Rayleigh number for the onset of convection. 

\subsection{Boundary Conditions}
In this paper we have used both rigid and free-slip velocity boundary conditions. The horizontal plates are assumed to be thermally conducting. For rigid and free-slip boundaries, the origin of the coordinate axes are taken at the mid plane and the bottom plate respectively. The positive $z$ axis are taken anti-parallel to the gravity.

Therefore, for rigid and thermally conducting boundaries we have 
\begin{eqnarray}
 u_x = u_y = u_z = \theta = 0 ~~~\mathrm{at} ~~~ z = \pm\frac{1}{2}, \label{bc1_r}
\end{eqnarray}
while, for free-slip conducting boundaries imply 

\begin{eqnarray}
{u_z} = \frac{\partial u_x}{\partial z}= \frac{\partial u_y}{\partial z} = \theta = 0~~~\mathrm{at} ~~~ z = 0, 1. \label{bc1_f}
\end{eqnarray}

Thus, the equations~(\ref{eq:momentum})-(\ref{eq:div_free}) together with the relevant boundary conditions provide the mathematical model of the rotating hydrodynamic system. 

\section{Linear Theory}

\begin{figure}[h]
\centering
\includegraphics[scale = .7]{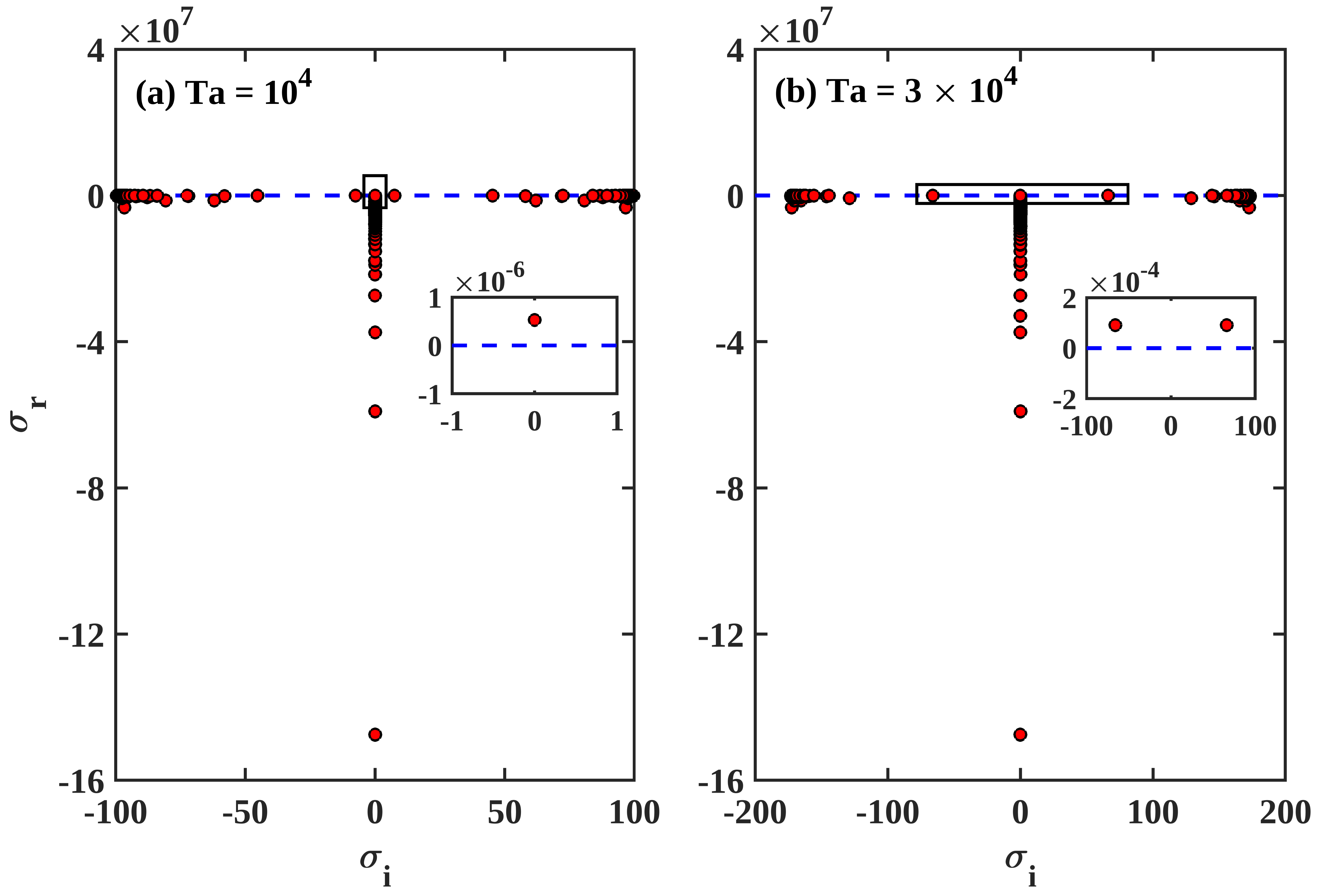}
\caption{Full eigenspectrum of the generalized eigen value problem (\ref{gev}) with $\mathrm{Pr} = 0.1$ in the presence of rigid boundary conditions just above the onset of convection for (a) stationary cellular convection and (b) overstable oscillatory convection.}
\label{eigen_spectrum}
\end{figure}
To determine the onset of convection, the linear stability analysis of the conduction state of the system is performed both with rigid and free-slip boundary conditions. First the convective fields are expanded in terms of the normal modes as
\begin{eqnarray}\label{normal mode}
\left[
    \begin{array}{c}
    {\bf u}    \\
      \theta   \\
       \pi \\
    \end{array}
   \right]=
   \left[
    \begin{array}{c}
    {\bf U}(z)    \\
      \Theta(z)  \\
       \Pi(z)  \\
    \end{array}
   \right] e^{i(k_x x+k_y y)+ \sigma t} + c.c., 
\end{eqnarray}
where $c.c.$, ${\bf U}(z) = \left(\mathrm{U}_x(z), \mathrm{U}_y(z), \mathrm{U}_z(z)\right)$, $k_x$ and $k_y$ stand for the  complex conjugate, the components of the wave vector in the $x$ and $y$ directions, respectively. The temporal growth rate of perturbations is represented by $\sigma = \sigma_r + i \sigma_i$. Substituting the above normal modes in the governing equations (\ref{eq:momentum}) - (\ref{eq:div_free}) and retaining the linear terms,  the following set of equations are obtained:
\begin{subequations}
\begin{gather}
(D^2-k^2-\sigma){\bf U}(z) - \sqrt{\mathrm{Ta}}({\bf U}(z)\times {\bf \hat{e}_3}) = ik_x \Pi(z) {\bf \hat{e}_1} \nonumber \\
+ ik_y \Pi(z) {\bf \hat{e}_2} +(D\Pi(z) - \mathrm{Ra} \Theta(z)) {\bf \hat{e}_3}, \\ 
(D^2 - k^2 - \sigma\mathrm{Pr}) \Theta(z) + \mathrm{U}_z(z) = 0 ,\\
ik_x \mathrm{U}_x(z) + ik_y \mathrm{U}_y(z) +D\mathrm{U}_z(z) = 0,
\end{gather}
\end{subequations}
where $D \equiv \frac{d}{dz}$ and $k = \sqrt{k_x^2 + k_y^2}$ is the horizontal wave number. These equations along with the considered boundary conditions are then discretized using a staggered-grid Chebyshev spectral collocation method~\cite{khorrami_report:1989} along the vertical direction as outlined in~\cite{mandal:POF_2022} which leads to a generalized eigenvalue problem given by

\begin{equation}
\mathrm{AX}=\sigma \mathrm{BX}, \label{gev}
\end{equation}
where, $\mathrm{A}$ and $\mathrm{B}$ are square matrices each of dimension $5N+4$, $N$ is the order of the Chebyshev polynomial, and $\sigma$ is the eigenvalue. Note that the matrices $\mathrm{A}$ and $\mathrm{B}$ are functions of the parameters of the system and the spatial grid points given by
\begin{equation}
\phi_l = \cos\left[\frac{\pi l}{N}\right]~(l=0,1,\dots, N),~\phi_{m+\frac{1}{2}} = \cos\left[\left(m + \frac{1}{2}\right)\frac{\pi}{N}\right]~(m=0,1,\dots, N-1).
\end{equation}
The vector $X$ is defined by
\begin{eqnarray*}
X =\left(\{{\mathrm U}_x(\phi_l)\}_{l=0}^N,\{{\mathrm U}_y(\phi_l)\}_{l=0}^N, \{{\mathrm U}_z(\phi_l)\}_{l=0}^N,\{{\Theta}(\phi_l)\}_{l=0}^N, \{\Pi(\phi_{m+\frac{1}{2}})\}_{m=0}^{N-1} \right)^T.
\end{eqnarray*}

\begin{figure}[h]
\centering
\includegraphics[scale = .55]{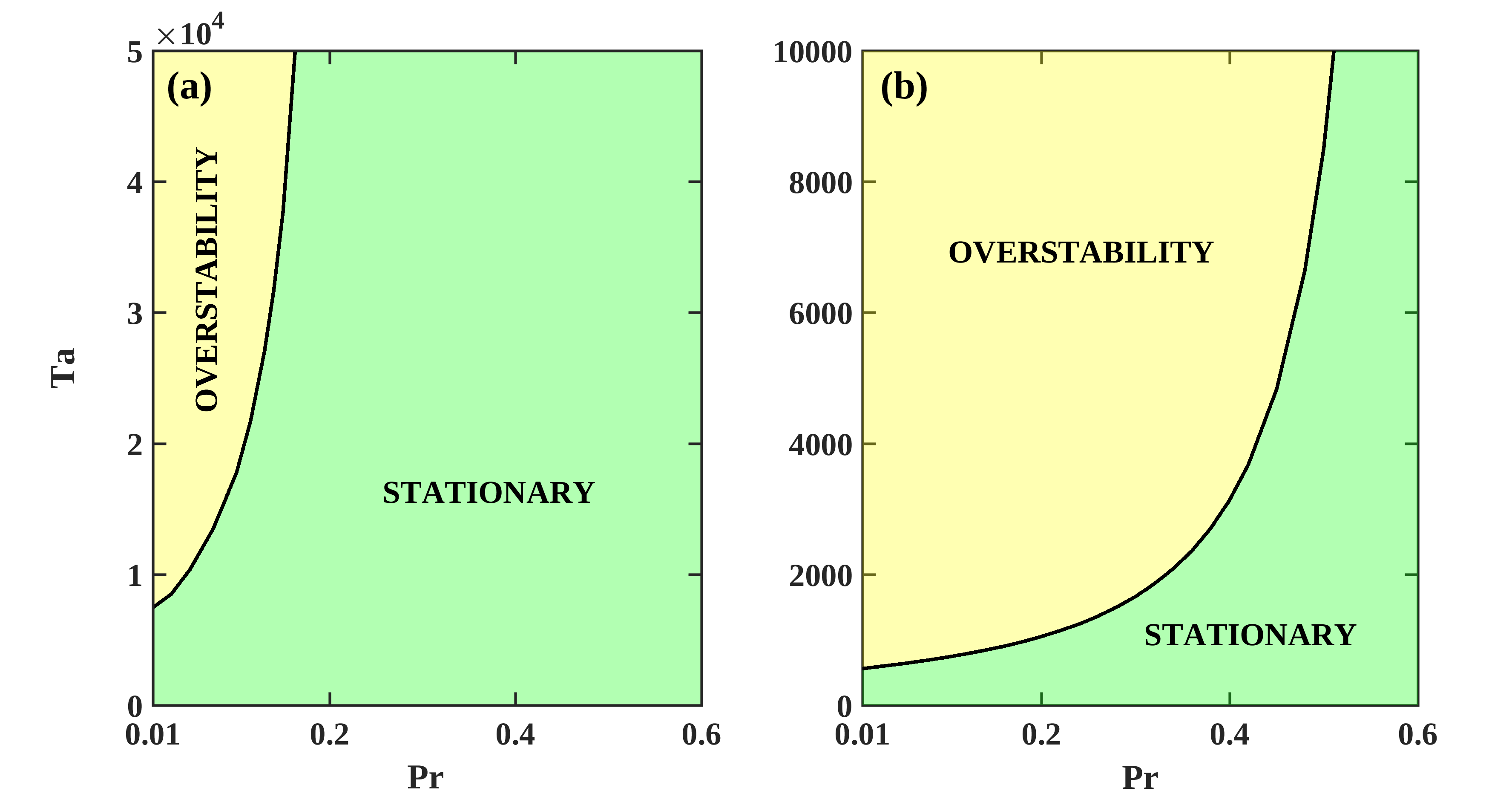}
\caption{Stationary and overstable flow regimes at the onset of convection on the $\mathrm{Pr} - \mathrm{Ta}$ plane for rigid (a) and free-slip (b) boundary conditions.}
\label{marginal_curve}
\end{figure}

Afterward, we proceed to solve the generalized eigenvalue problem utilizing the QZ algorithm.
The trivial conduction state becomes unstable when the real part of one of the eigen values become positive from negative with the variation of a relevant parameter. One obtains the case of stationary cellular convection, where the so called `principle of exchange of stability'~\cite{chandra:book} is valid, when a real eigen value of the problem becomes positive. On the other hand, the case of overstability is obtained when the real parts of a pair of complex conjugate eigen values become positive.  Figure~\ref{eigen_spectrum} show the typical eigen spectrum of the generalized eigen value problem for rigid boundary conditions for two specific sets of parameter values.
 We then solve the generalized eigen value problem in the entire region of the parameter space considered in this paper to determine the critical Rayleigh number $\mathrm{Ra_c}$ and the critical wave number $k_c$ for the onset of convection. The critical wave numbers determined from the linear theory are then used subsequently for low dimensional modeling and performing direct numerical simulations. Figures~\ref{marginal_curve}(a)  and (b) show the regions of stationary and overstable convection regimes on the $\mathrm{Pr} - \mathrm{Ta}$ plane for 
rigid and free-slip boundary conditions respectively, determined from the linear theory. 

\begin{table*}
\caption{Critical Rayleigh numbers ($\mathrm{Ra_c}$) for different $\mathrm{Ta}$, computed from the linear theory (LT), 1D model and DNS for $\mathrm{Pr}=0.1$ with rigid boundary conditions.}\label{table:comparison_of_Rac}
\begin{tabularx}{\textwidth} { 
   >{\centering\arraybackslash}X 
   >{\centering\arraybackslash}X 
   >{\centering\arraybackslash}X
   >{\centering\arraybackslash}X
   >{\centering\arraybackslash}X
   >{\centering\arraybackslash}X
   >{\centering\arraybackslash}X
   >{\centering\arraybackslash}X }
 \hline
 \hline
 $\mathrm{Ta}$ & $\mathrm{k_c}$& $\mathrm{Ra_c}$  & $\mathrm{Ra_c}$ & $\mathrm{Ra}_c$ & Error(\%) & Error(\%) \\
               & (LT)  & (LT) & (Model) & (DNS) & (LT vs Model) & (LT vs DNS)\\
\hline
$1\times 10$   & $3.119$  & $1720$   &  $1734$ & $1715$        &       $0.81$ &  $0.29$   \\
$1\times 10^2$ & $3.159$  & $1764$   &  $1779$ & $1760$       &       $0.85$ &   $0.23$ \\
$5\times 10^2$ & $3.317$  & $1948$   &  $1976$ & $1940$        &       $1.43$ &   $0.41$   \\
$1\times 10^3$ & $3.482$  & $2159$   &  $2205$ & $2150$       &       $2.13$ &   $0.42$   \\
$5\times 10^3$ & $4.263$  & $3476$   &  $3706$ & $3380$        &       $6.61$ &   $2.76$   \\
\hline
\hline
\end{tabularx}
\end{table*}

\section{Nonlinear Analysis}
Linear theory only determines the marginally stable state of the system, but it can not determine the flow patterns at the onset which is purely a nonlinear phenomenon. Thus, to investigate the flow patterns at the onset of convection we perform nonlinear analysis of the system. For the nonlinear analysis we employ low dimensional modeling technique 
and simultaneously perform direct numerical simulations of the system both for rigid and free-slip boundary conditions. Results are discussed below. 

\subsection{Stationary cellular convection}
The primary objective of the paper is to investigate the onset of rotating stationary cellular convection and the related finite amplitude solution, where the principle of exchange of stability is valid. The detailed results obtained for rigid as well as free-slip boundary conditions are presented subsequently. 
\subsubsection{Results for rigid boundary conditions}
Here we consider rigid velocity boundary conditions for the analysis of finite amplitude two dimensional rolls solution of subcritical origin at the onset of convection. First we perform low dimensional modeling of the system followed by the direct numerical simulations. A minimal mode low dimensional model is constructed by considering the following  truncated expansions
\begin{eqnarray}
u_z (x,y,z,t) &=& W_{101}(t)\mathrm{cos}(k_cx) C_1(\lambda_1 z), \nonumber \\
\omega_z(x,y,z,t) &=& Z_{201}(t) \mathrm{cos}(2k_cx)\mathrm{cos}(\pi z) 
                  + Z_{102}(t)\mathrm{cos}(k_cx)\mathrm{sin}(2\pi z), \nonumber \\
\theta (x,y,z,t) &=& T_{101}(t)\mathrm{cos}(k_cx)\mathrm{cos}(\pi z) 
                  + T_{002}(t) \mathrm{sin}(2\pi z), \nonumber
\end{eqnarray} 
of vertical velocity $u_z$,  vertical vorticity $\omega_z$ and temperature $\theta$ respectively, in terms of the Chandrasekhar function~\cite{chandra:book} $C_1(\lambda_1 z) = \frac{\cosh\lambda_1 z}{\cosh \lambda_1/2}-\frac{\cos \lambda_1 z}{\cos \lambda_1/2}$ with $\lambda_1 = 4.73$, $\sin$ and $\cos$ functions compatible with the boundary conditions. The experimental~\cite{rossby:JFM_1969} and numerical~\cite{clever1:JFM_1979,mandal:POF_2022} observation of two dimensional rolls solutions at the onset of rotating convection, leads to the natural choice of the perturbations $W_{101}(t)\mathrm{cos}(k_cx) C_1(\lambda_1 z)$ in $u_z$, $ \omega_z = 0$ and $T_{101}(t)\mathrm{cos}(k_cx)\mathrm{cos}(\pi z)$ with respect to which the system is marginally stable at the critical Rayleigh number. Next, for the saturation of the marginally stable mode of convection in a simplest possible way, we consider modes $T_{002}(t) \mathrm{sin}(2\pi z)$ in the temperature, and $Z_{201}(t) \mathrm{cos}(2k_cx)\mathrm{cos}(\pi z)$ and $Z_{102}(t)\mathrm{cos}(k_cx)\mathrm{sin}(2\pi z)$ in vertical vorticity. 

Now projecting the governing hydrodynamics equations on these five modes, the following set of coupled nonlinear ordinary differential equations is obtained:
\begin{eqnarray}
\dot{\xi} &=& (a_{11} \xi + a_{12} \zeta  +  a_{13} \phi)/a_{14}, \label{5eq1}\\
\dot{\zeta}&=& a_{21} \zeta + a_{22} \xi  + a_{23} \eta \xi, \\
\dot{\eta} &=&  - a_{31} \eta + a_{32}\zeta\xi, \\
\dot{\phi} &=& (a_{41}\phi + a_{42} \xi + a_{43} \psi \xi)/\mathrm{Pr},\\
\dot{\psi} &=& (- a_{51} \psi + a_{52}  \xi \phi )/\mathrm{Pr}, \label{5eq5}
\end{eqnarray} 
where $\xi = W_{101} $, $\zeta = Z_{102}$, $\eta = Z_{201} $, $\phi = T_{101}$ and $\psi = T_{002}$, $a_{11}=-(19.74 k^4+9880.87 + 485.70 k^2),$  $a_{12}=48.75 \sqrt{\mathrm{Ta}}$, $a_{13}=13.76 \mathrm{Ra}k^2,$ $a_{14}=19.74(k^2+12.30),$ $a_{21}=-(39.47+k^2), a_{22}=-4.93 \sqrt{\mathrm{Ta}}$, $a_{23}= 2.09$, $a_{31}=9.86 + 4k^2, a_{32}=-8.36$, $a_{41}=-(0.99 k^2 + 9.86),$ $a_{42}=1.39,$ $a_{43}=-5.09  \mathrm{Pr},$ $a_{51}=39.47, ~\mathrm{and}~a_{52}=2.54 \mathrm{Pr}.$

For the validation of the above model, we first determine the onset of stationary convection from model (\ref{5eq1}) - (\ref{5eq5}) and compare with the linear theory results. The comparisons are shown in the table~\ref{table:comparison_of_Rac}. From the table, a satisfactory match between the linear theory and the model results for the onset of convection is clear. Next we move ahead to reduce the model further to achieve the simplest possible description of the system for the onset of convection in the stationary regime using adiabatic elimination process~\cite{glendinning:2001}.

To investigate the dynamics of the dynamical system (\ref{5eq1}) - (\ref{5eq5}) beyond the onset of convection, we first focus on the stationary convection regime and for that we need to determine the fixed points of it in the parameter regime of our interest. Thus, we obtain a set of $5$ algebraic equations in $\xi, \zeta, \eta, \phi$ and $\psi$ by equating the right hand sides of the equations (\ref{5eq1}) - (\ref{5eq5}). Next, from those $5$  equations we eliminate the variables $\zeta, \eta, \phi$ and $\psi$, and obtain the following equation  in $\xi$:
\begin{equation}
c_1\xi+c_3\xi^3+c_5\xi^5=0, \label{eq:1Dfinal_eq}
\end{equation}  
where
$c_5=a_{23}a_{11}a_{32}a_{43}a_{52}, c_3=a_{11}(a_{21}a_{31}a_{43}a_{52}+a_{23}a_{32}a_{41}a_{51})- (a_{12}a_{31}a_{22}a_{43}a_{52}+a_{13}a_{42}a_{51}a_{23}a_{32}),
c_1 = a_{51}a_{31}[a_{11}a_{41}a_{21} - (a_{12}a_{22}a_{41}+a_{13}a_{42}a_{21})].$
Now using the equation (\ref{eq:1Dfinal_eq}) we construct the following potential function~\cite{verma:POP_2013,stogatz:book} for $\xi$:
\begin{equation}
V(\xi) = -\int {(c_1\xi + c_3\xi^3 + c_5\xi^5)}d\xi. \label{5eq7}
\end{equation}

Note that, although, the potential function $V(\xi)$ defined in equation (\ref{5eq7}) contains the variable $\xi$ only, the contributions of the other variables are embedded there in the coefficients. Subsequently, it can be seen that $V(\xi)$ nicely captures the phenomenon of transition to subcritical convection.  Now using the potential function $V(\xi)$ we obtain the following one dimensional dynamical system
\begin{equation}
\dot{\xi} = - \frac{dV({\xi})}{d\xi} = c_1\xi + c_3\xi^3 + c_5\xi^5, \label{1deq}
\end{equation}
which is used subsequently to investigate the onset of subcritical and supercritical onset of rotating convection.
Interestingly, the one dimensional dynamical system (\ref{1deq}) captures the dynamics of the $5$ model [(\ref{5eq1}) - (\ref{5eq5})] in the stationary convection regime very closely. The onset of stationary convection determined from the 1D model is same as the ones presented in the table~\ref{table:comparison_of_Rac}. We now perform detailed bifurcation analysis of the model (\ref{1deq}) using an open source software XPPAUT~\cite{ermentrout_XPP:book} to understand the transition to convection for different values of the parameters. 
\begin{figure}[h]
\centering
\includegraphics[scale = .7]{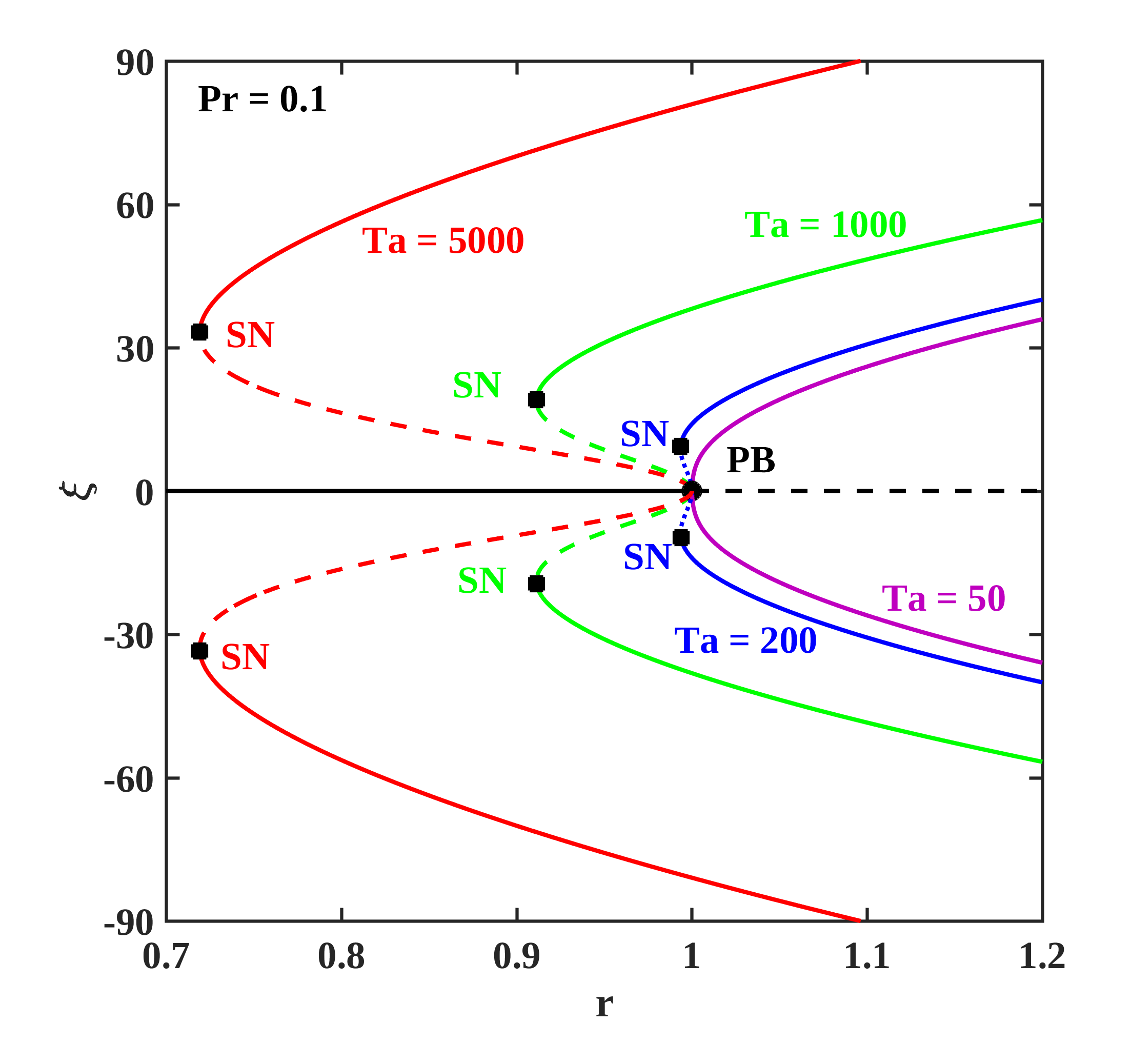}
\caption{Bifurcation diagrams prepared from the 1D model for fixed $\mathrm{Pr} = 0.1$ and four different $\mathrm{Ta}$. Pitchfork and saddle-node bifurcation points are marked with PB and SN respectively. Supercritical pitchfork bifurcation is seen for $\mathrm{Ta} = 50$, while, subcritical pitchfork bifurcations are observed for other three $\mathrm{Ta}~(200, 1000, 5000)$.}
\label{bifurcation_Pr0p1}
\end{figure}

\begin{figure}[h]
\centering
\includegraphics[scale = .7]{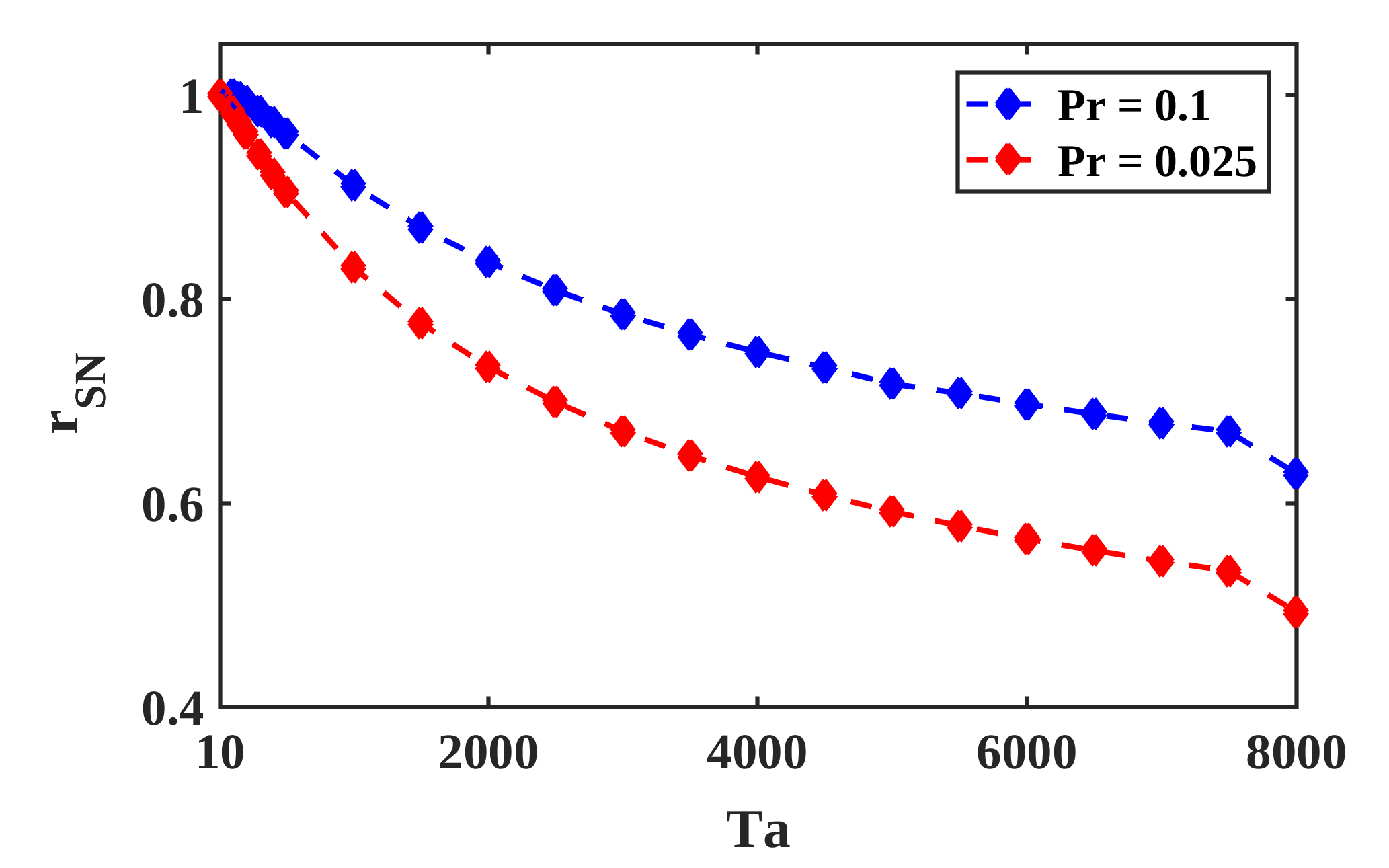}
\caption{Variation of the saddle node bifurcation point with $\mathrm{Ta}$ for two different $\mathrm{Pr}$.}
\label{saddlenode_point}
\end{figure}

\begin{figure}[h]
\centering
\includegraphics[scale = .7]{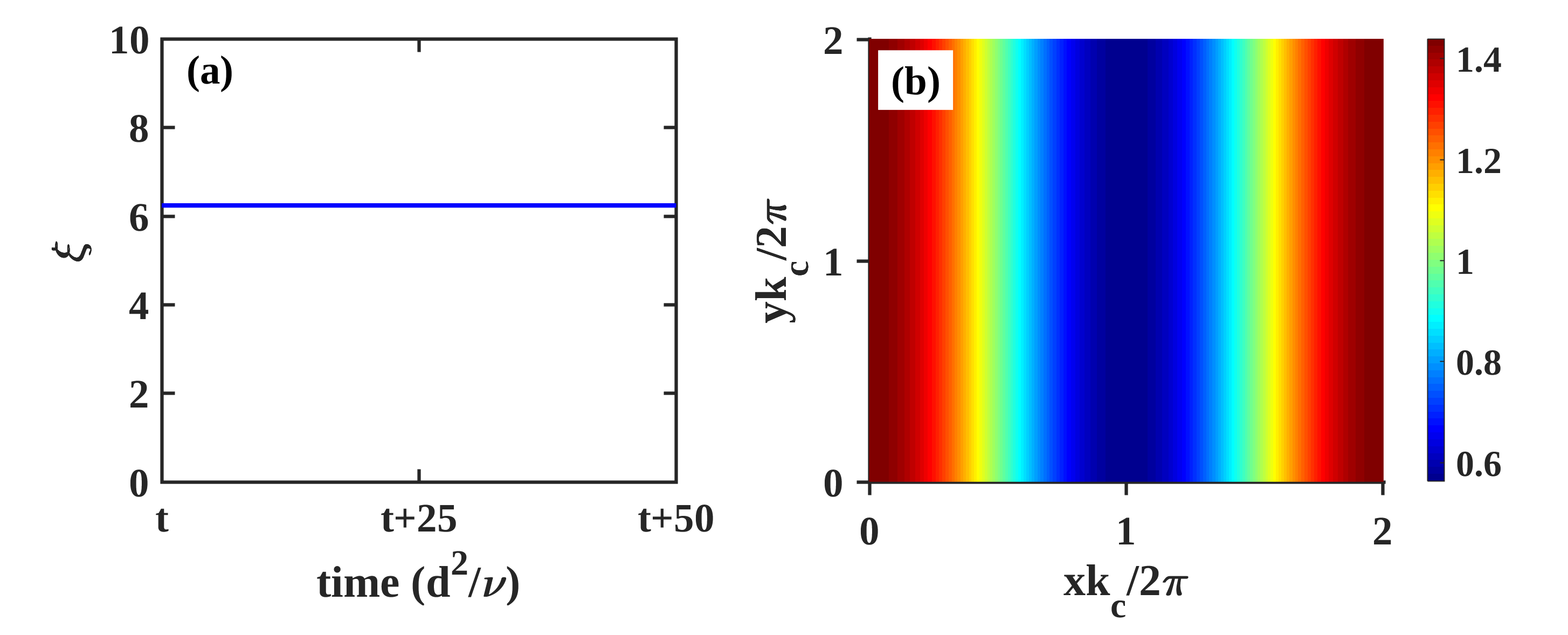}
\caption{Time series (a) and corresponding flow pattern (b) at the onset of convection for $\mathrm{Ta}=50$ and $\mathrm{Pr}=0.1$. }
\label{timeseries}
\end{figure}

Figure~\ref{bifurcation_Pr0p1} shows the bifurcation diagrams constructed from the 1D model for fixed $\mathrm{Pr} = 0.1$ and various values of $\mathrm{Ta}$. In the bifurcation diagrams, for each values of $\mathrm{Ta}$, the variation of $r$ is shown along the horizontal axis and that of the stable (solid lines) and unstable (dashed lines) fixed points along the vertical axis. The bifurcation diagrams show transition to convection through supercritical and subcritical pitchfork bifurcations for slow and high rotation rates respectively.  Note that the subcritical transition to convection is characterized by the appearance of a saddle-node (SN) bifurcation for higher $\mathrm{Ta}$ which results in a discontinuity in the solutions followed by finite amplitude flow patterns at the onset.  It is also observed that as the rotation rate is increased, the saddle-node (SN) bifurcation point moves towards lower $r$ and the distance between the pitchfork bifurcation (PB) and SN points increase. For more details, we look at the variation of the location of the SN bifurcation point with $\mathrm{Ta}$ for $\mathrm{Pr} = 0.025$ and $0.1$ and the results are presented in the figure~\ref{saddlenode_point}. Form the above discussion, it is apparent that for a fixed $\mathrm{Pr}$, the rotation promotes subcriticality.
The time series of $\xi$ and fluid pattern associated with the stable stationary solutions of supercritical origin is shown in the figure~\ref{timeseries}. The flow patterns at the onset of subcritical convection is also similar but the associated mean velocity is higher.  

\begin{figure}[h]
\centering
\includegraphics[scale = .52]{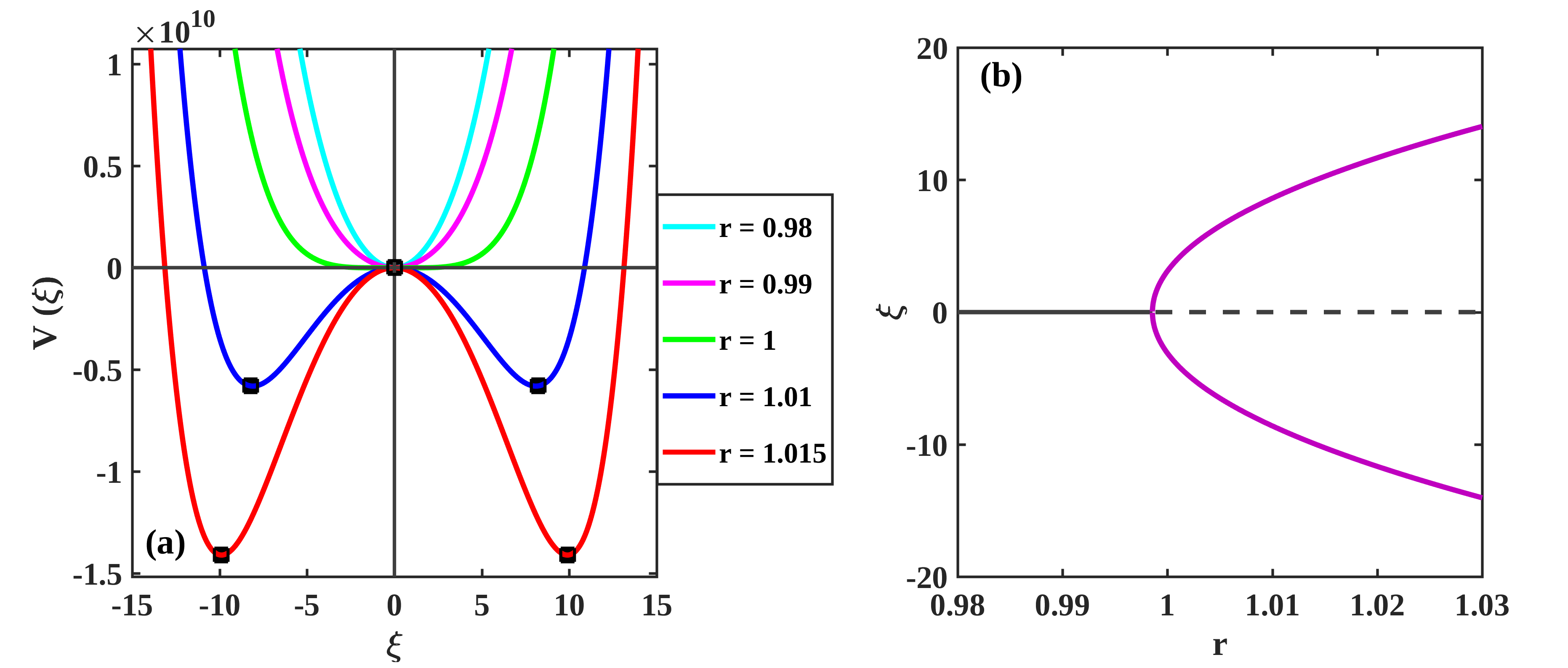}
\caption{{ Supercritical transition to convection for $\mathrm{Pr} = 0.1$ and $\mathrm{Ta} = 10$. (a) Graphs of $V(\xi)$ for different values of $r$ around the critical point $r=1$. The extremum points are shown with black squares. (b) Bifurcation diagram of the 1D model (\ref{1deq}) showing supercritical pitchfork bifurcation.}}
\label{potential_supercritical}
\end{figure}

To illustrate the transitions from supercritical to subcritical onset of convection with the variation of rotation rate ($\mathrm{Ta}$) in more detail, we consider two values of $\mathrm{Ta}$, namely, $10$ and $500$ where, respectively, supercritical and subcritical transition to convection are observed in direct numerical simulations~\cite{clever1:JFM_1979,mandal:POF_2022} with $\mathrm{Pr} = 0.1$. For $\mathrm{Ta} = 10$, we first draw the graphs of the potential function $V(\xi)$ by varying the reduced Rayleigh number $r$ around $r = 1$, the critical Rayleigh number for the onset of convection. The graphs are shown in the figure~\ref{potential_supercritical}(a).  

From the figure~\ref{potential_supercritical}(a), only single well potentials are observed for $0< r \leq 1$. The shape of the graphs suggest that $\xi = 0$ is the only fixed point of the system and it is stable, which physically indicates the stability of the conduction state of the system. On the other hand, for $r > 1$, the shape of the graphs change and double well potentials are observed. Now, there are three different fixed points (extrema) of the system, namely, the trivial $\xi=0$ and two other non-zero fixed points (say $\pm \xi^*$) which are marked with black squares in the figure~\ref{potential_supercritical}(a). The shape of the graphs indicate that the trivial fixed point is unstable (maxima) and the non-zero fixed points are stable (minima). Note that the stable non-zero fixed points physically represent the stationary two dimensional rolls patterns. Such a scenario typically observed around a supercritical pitchfork bifurcation~\cite{stogatz:book}. The bifurcation diagram of the one dimensional model (\ref{1deq}) for the same set of parameter values is shown in the figure~\ref{potential_supercritical}(b) confirm the supercritical nature of the transition to convection. 
\begin{figure}[h]
\centering
\includegraphics[scale =.55]{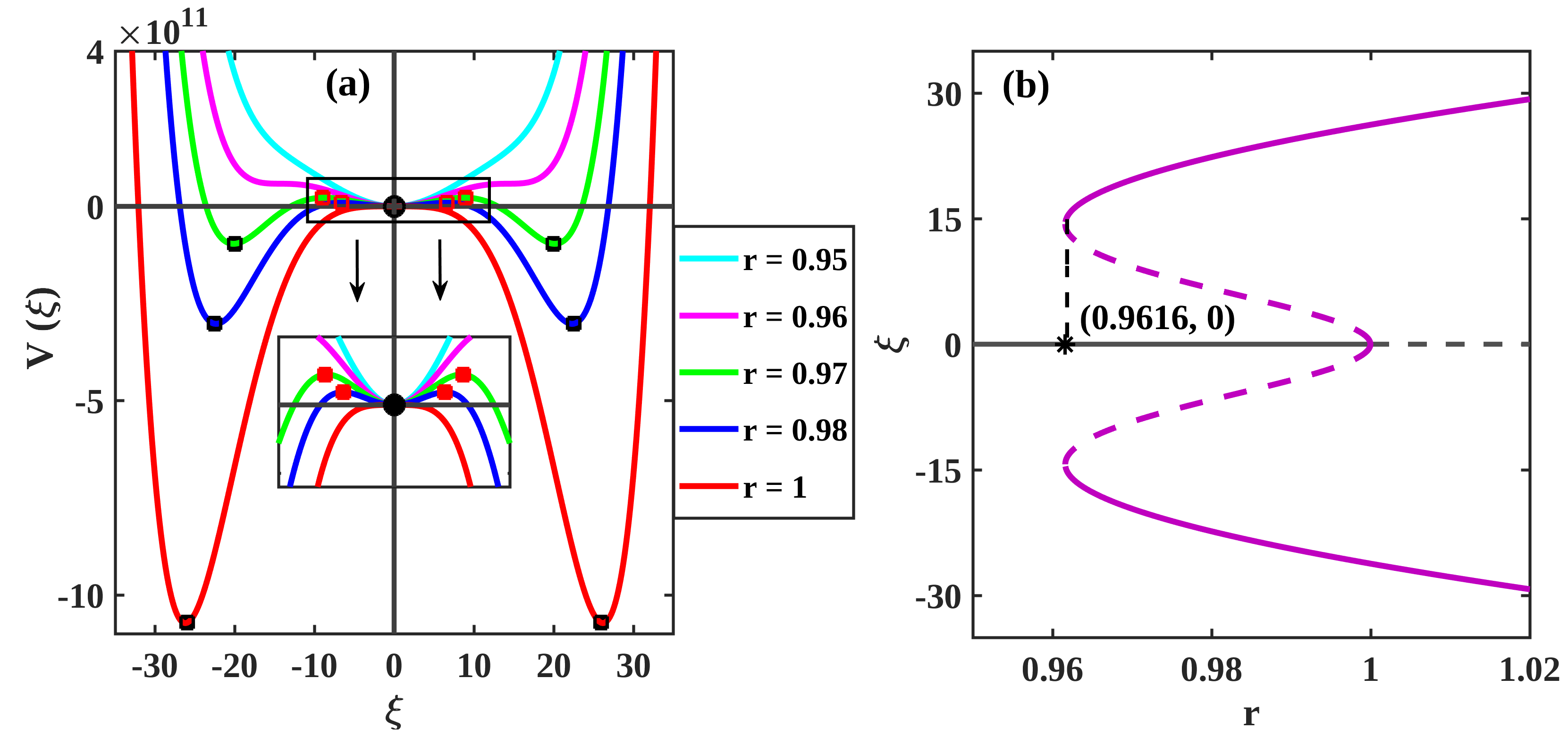}
\caption{Subcritical transition to convection for $\mathrm{Pr} = 0.1$ and $\mathrm{Ta} = 500$. (a) Graphs of $V(\xi)$ for different values of $r$ around the critical point $r=1$. The filled black circle, red and black squares show the locations of the extremum points. (b) Bifurcation diagram of the 1D model (\ref{1deq}) showing subcritical pitchfork bifurcation.}
\label{potential_subcritical}
\end{figure}

Next, for $\mathrm{Pr} = 0.1$ and $\mathrm{Ta} = 500$, we also draw the graphs of $V(\xi)$ by varying $r$ around the critical value $r =1$ (figure~\ref{potential_subcritical} (a)). The fixed points of the system can be identified from the location of the local maxima or minima points of the graph marked by filled black circle, red and black squares respectively. The local maxima and minima respectively represent unstable and stable fixed points of the system. In this case, only the trivial fixed point ($\xi = 0$) exists and stable (filled black circle in the figure~\ref{potential_subcritical} (a)) for $0 < r < 0.9616$. Interestingly, five different fixed points (three stable and two unstable) exist in the range $ 0.9616 < r < 1$. The stable fixed points (filled black circle and squares) are separated by the unstable fixed points (red squares).   As a result, the phenomenon of hysteresis is observed in this range of $r$. Further increase of $r$ beyond $r=1$, two unstable non-zero fixed points ceased to exist and only three fixed points exist. The trivial fixed point become unstable and finite amplitude stable non-zero fixed points continue to exist. Thus, at the onset of convection, finite amplitude two dimensional steady flow patterns are observed just at the onset of convection ($r > 1$).  This is a signature of subcritical pitchfork bifurcation with hysteresis~\cite{stogatz:book} and related transition to finite amplitude convection is subcritical in nature for the considered set of parameter values. Finally, we use the one dimensional model once again to construct a bifurcation diagram  for the parameter values $\mathrm{Pr} = 0.1$ and $\mathrm{Ta} = 500$. The bifurcation diagram is shown in the figure~\ref{potential_subcritical} (b). The bifurcation clearly shows the scenario of subcritical pitchfork bifurcation with hysteresis around the critical point $r = 1$. 
\begin{figure}[h]
\centering
\includegraphics[scale = .7]{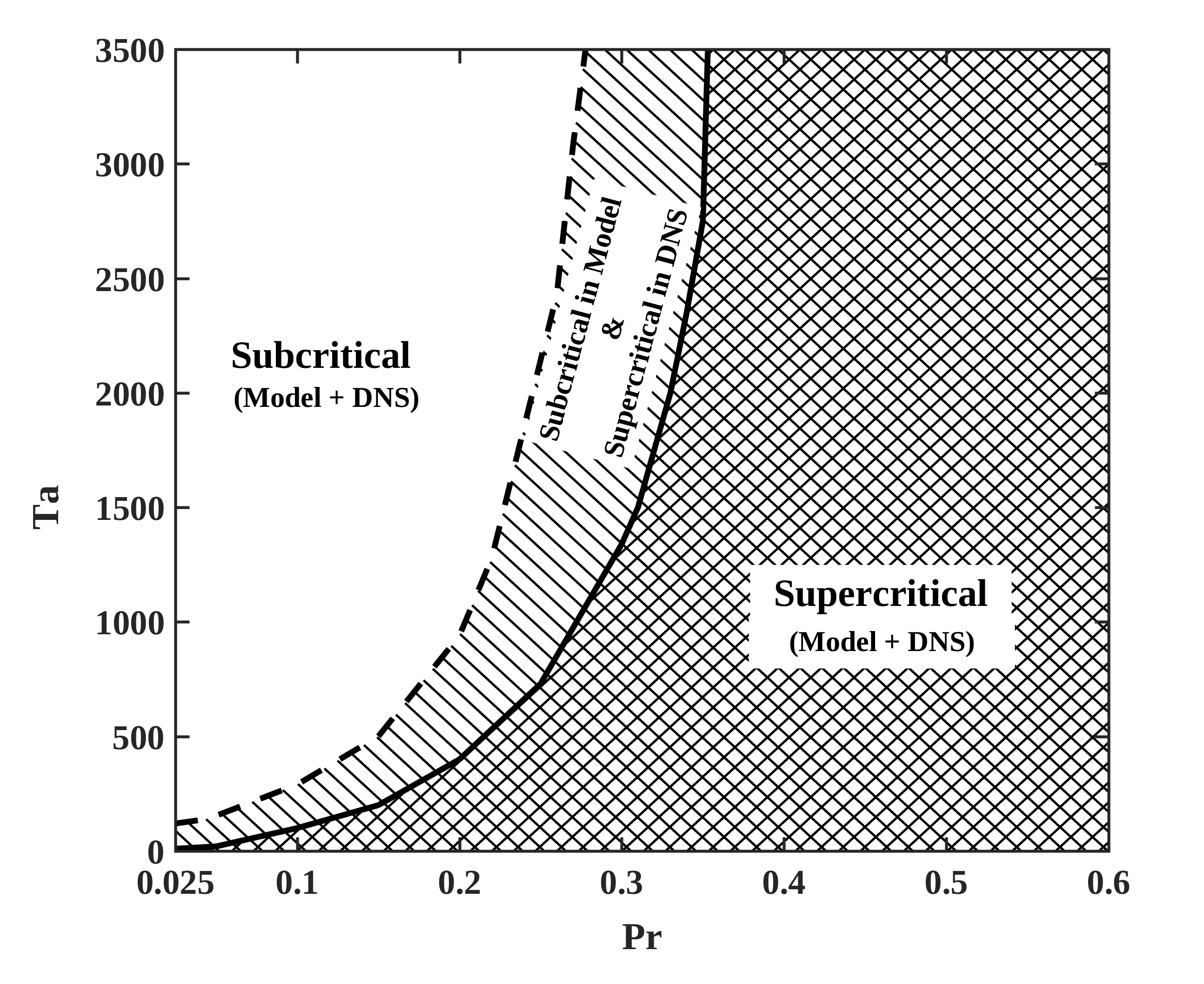}
\caption{Two parameter diagram  constructed from the 1D model (\ref{1deq}) and DNS for rigid boundary conditions demarcating supercritical and subcritical flow regimes on the $\mathrm{Pr} -\mathrm{Ta}$ plane. The thick solid and dashed black curves are obtained from the model and DNS respectively. }
\label{supercritical_subcritical}
\end{figure}

Therefore, from the foregoing analysis of the one dimensional model and the potential function $V(\xi)$, it is seen that the change of shape of the graphs of the the potential function  with the variation of the parameter $r$ determines the nature of transition to convection. The potential function, although, is a highly simplified description of the system under consideration, yet, it is able to capture the transition phenomena quite satisfactorily. Interesting to note hare that the mode $Z_{201}$ considered in the vertical vorticity is very important in capturing the subcritical behavior in the system. We did not get subcritical convection excluding this particular mode, even by considering large number of modes in the low dimensional modeling. We now use the 1D model to demarcate the supercritical and subcritical onset of convection on the $\mathrm{Pr} - \mathrm{Ta}$ plane and results are shown in the figure~\ref{supercritical_subcritical}. It is clearly observed that the rotation rate ($\mathrm{Ta}$) promotes subcriticality, while the Prandtl number ($\mathrm{Pr}$) inhibits it by promoting supercriticality. 

Now, to check the validity of the model results we perform three dimensional direct numerical simulations of the system using the open source spectral element code NEK5000~\cite{fischer2008} in a rectangular box of size $\frac{2\pi}{k_c} \times \frac{2\pi}{k_c} \times 1$ with grid resolution $56 \times 56 \times 56$. Time advancement is done by a suitable second order backward difference scheme with Courant-Friedrichs-Lewy (CFL) condition in the code. We use random initial conditions with time step $\delta t = 1\times 10^{-4}$ for all the simulations. At the outset, we use the critical wave number $k_c$ computed from the linear theory and determine the critical Rayleigh number for the onset of convection. The results for $\mathrm{Pr} = 0.1$ are presented in the table~\ref{table:comparison_of_Rac} which show a satisfactory match among the linear theory, model and DNS.

\begin{figure}[h]
\centering
\includegraphics[scale = .7]{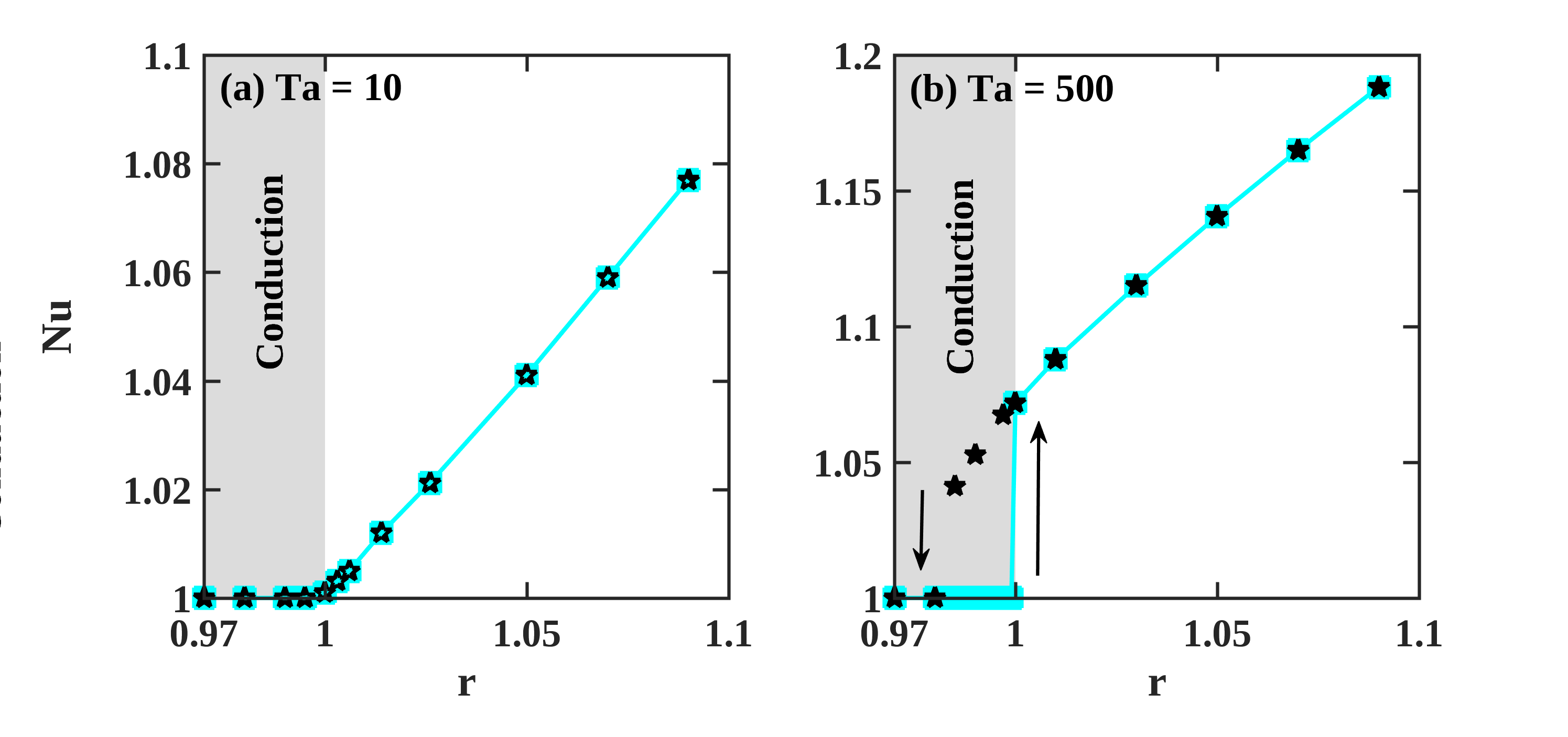}
\caption{Variation of the Nusselt number $\mathrm{Nu}$ with $\mathrm{r}$ for $\mathrm{Pr} = 0.1$. The solid cyan line and black stars respectively correspond to the forward and backward continuation data for (a) $\mathrm{Ta} = 10$ and (b) $\mathrm{Ta} = 500$.}
\label{NS_nusselt_number}
\end{figure}

In DNS, the subcritical and supercritical transitions to rotating convection are determined by computing the Nusselt number $\mathrm{Nu}$, defined by
\begin{equation}
\mathrm{Nu} = 1 + \mathrm{Pr}^2\langle u_z\theta\rangle, 
\end{equation}
which measures ratio of the average convective to conductive heat transfers across the layers. The symbol $\langle \cdot \rangle$ represents the spatial average over the computational domain. 
Supercritical or subcritical convection occurs at the onset if the value of $\mathrm{Nu}$ follows same or different paths during the the forward and backward continuation of the reduced Rayleigh number across the critical point $r =1.$ 
Figure~\ref{NS_nusselt_number} shows the variations of the Nusselt numbers for $\mathrm{Ta} = 10$ and $500$. From the figure~\ref{NS_nusselt_number}(a), it is observed that for forward as well as backward continuation of $r$, the Nusselt number follow the same path and $\mathrm{Nu}$ does not show any jump at the critical point $r=1$ indicating supercritical nature of transition. On the other hand, figure~\ref{NS_nusselt_number}(b) shows a discontinuity in $\mathrm{Nu}$, and forward and backward data following different paths forming the so called `hysteresis loop' indicating the subcritical transition to convection. For detailed understanding about the parameter space, we now determine the curve separating the supercritical and subcritical onset of convection flow regime which is shown using the thick dashed black curve in the figure~\ref{supercritical_subcritical}. Thus, the model and the DNS results show qualitatively similar behavior. Subsequently, we investigate the phenomenon of subcritical convection in RRBC in the presence of free-slip boundary conditions.  

\subsubsection{Results for free-slip boundary conditions}
It is interesting to note that subcritical convection was first reported in RRBC model in the presence of free-slip boundary conditions~\cite{veronis:1959,veronis:1966} before the experimental observation of the phenomenon by Rossby~\cite{rossby:JFM_1969}. Both weakly nonlinear theory\cite{veronis:1959} and low dimensional modeling\cite{veronis:1966} was used for the investigation. Following the low dimensional modeling approach presented in~\cite{veronis:1966}, we expand the convective vertical velocity, vorticity and temperature fields as follows: 
\begin{eqnarray}
v_z  &=& W_{101}(t)\cos{kx}\sin{\pi z},\\
\omega_z &=& Z_{101}(t)\cos{kx}\cos{\pi z} + Z_{200}\cos{2\pi x},\\
\theta &=& T_{101}(t)\cos{kx}\sin{\pi z} + T_{002}(t)\sin{2\pi z},
\end{eqnarray}
in terms of the boundary condition compatible basis functions. We then project the hydrodynamic equations (\ref{eq:momentum}) - (\ref{eq:div_free}) on these modes to obtain a five dimensional coupled ordinary differential equations for the Fourier amplitudes $W_{101}$, $Z_{101}$, $Z_{200}$, $T_{101}$ and $T_{002}$ which is given by 
\begin{eqnarray}
\dot{\xi}&=& -(a^2 \xi -c \phi +b \gamma)/a \label{5m_f1}\\
\dot{\gamma}&=& -a \gamma +b \xi-\pi/2 \chi \xi  \\
\dot{\chi} &=& -4 k^2 \chi+\pi \xi \gamma  \\
\dot{\phi}&=& -(a \phi -\xi-\pi \mathrm{Pr}\xi \psi)/\mathrm{Pr}  \\
\dot{\psi}&=& -\pi/2(8\pi \psi +\mathrm{Pr} \xi \phi)/\mathrm{Pr}  \label{5m_f5}
\end{eqnarray} 
where $ \xi=W_{101},~\gamma=Z_{101},~\chi=Z_{200},~\phi=T_{101},~\psi=T_{002},~a=\pi^2+k^2,~b=\pi \sqrt{\mathrm{Ta}},~c=\mathrm{Ra}k^2$.

As done with rigid boundary conditions, here also we compare the critical Rayleigh number for the onset of convection determined from the linear theory (LT) and the above model. The comparison results are presented in the table~\ref{table:comparison_of_Rac_FS} which shows a very good match.
\begin{table}
\caption{Critical Rayleigh numbers ($\mathrm{Ra_c}$) for different $\mathrm{Ta}$ computed from linear theory (LT), 1D model and DNS for $\mathrm{Pr}=0.1$ with free-slip boundary conditions.}\label{table:comparison_of_Rac_FS}
\begin{tabularx}{\textwidth} { 
   >{\centering\arraybackslash}X 
   >{\centering\arraybackslash}X 
   >{\centering\arraybackslash}X
   >{\centering\arraybackslash}X
   >{\centering\arraybackslash}X
   >{\centering\arraybackslash}X
   >{\centering\arraybackslash}X
   >{\centering\arraybackslash}X }
 \hline
 \hline
 $\mathrm{Ta}$ & $\mathrm{k_c}$& $\mathrm{Ra_c}$  & $\mathrm{Ra_c}$ & $\mathrm{Ra}_c$ & Error(\%) & Error(\%) \\
               & (LT)  & (LT) & (Model) & (DNS) & (LT vs Model) & (LT vs DNS)\\
\hline
$1$   & $2.226$  & $659.5$   &  $659.5$ & $660$        &       $0$ &  $0.07$   \\
$10$ & $2.269$  & $677.1$   &  $677.1$ & $675$       &       $0$ &   $0.31$ \\
$50$ & $2.434$  & $748.3$   &  $648.3$ & $720$        &       $0$ &   $3.78$   \\
$100$ & $2.594$  & $826.2$   &  $826.2$ & $775$       &       $0$ &   $6.19$   \\
$500$ & $3.277$  & $1274.6$   &  $1274.5$ & $1090$        &       $0.01$ &   $14.48$   \\
\hline
\hline
\end{tabularx}
\end{table}
We then further reduce the set of five ordinary differential equations to a single one by adopting the similar procedure as described in the previous subsection. Thus, we obtain the following one dimensional model
\begin{equation}
\dot{\xi} =  d_1\xi + d_3\xi^3 + d_5\xi^5, \label{1deq_f}
\end{equation}
for the investigation of subcritical bifurcation, where,
$\xi = W_{101}$, $d_1=64a^4(a-\pi^2)+64ab^2(a-\pi^2)-64ac(a-\pi^2)$, $d_5=\pi^2 \mathrm{Pr}^2a^2$, $d_3=8a^3\mathrm{Pr}^2(a-\pi^2)+8a^3\pi^2+8b^2\mathrm{Pr}^2(a-\pi^2)-8c\pi^2$.

\begin{figure}
\centering
\includegraphics[scale = 0.7]{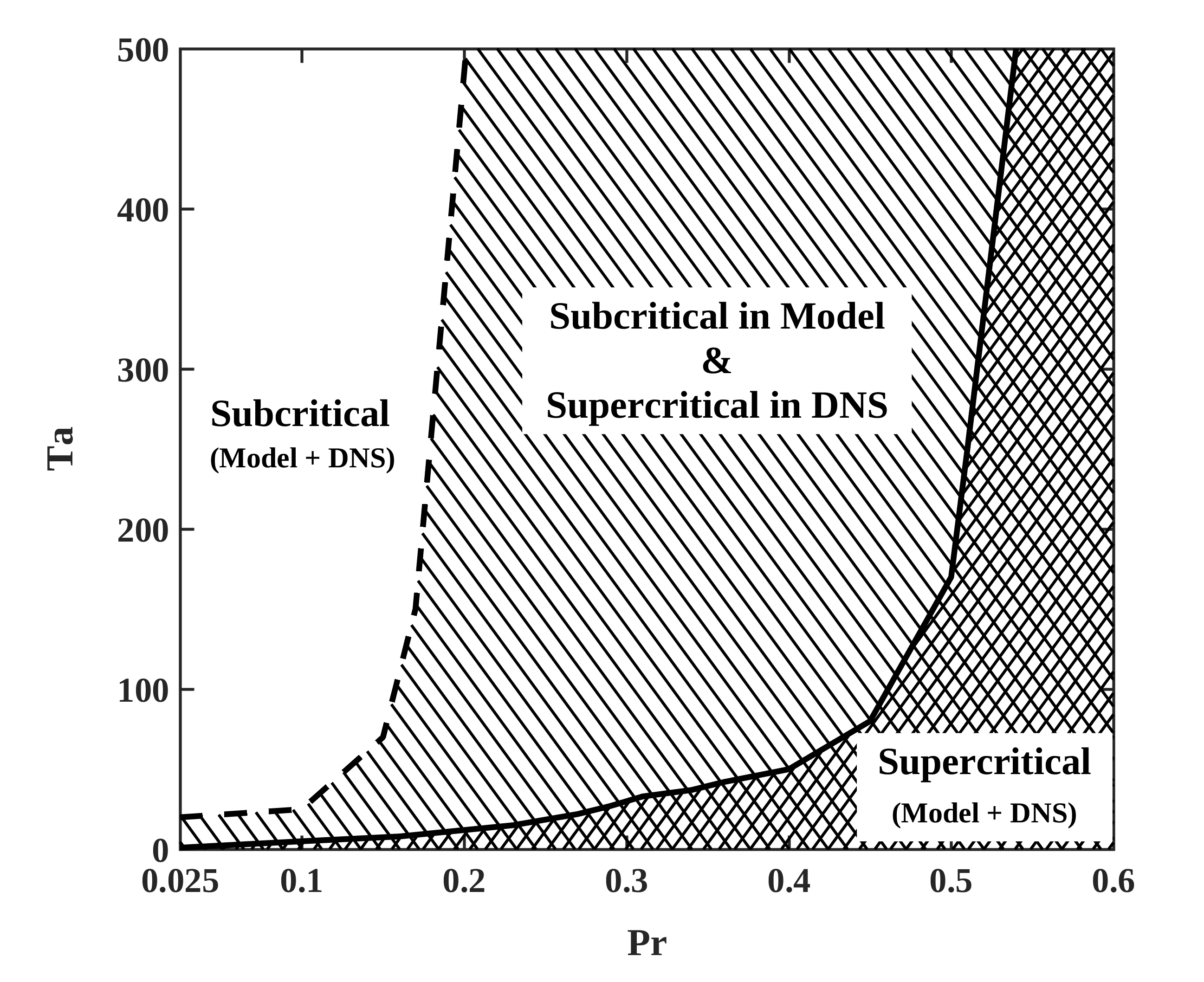}
\caption{Two parameter diagram  computed from the 1D model and DNS for free-slip boundary conditions demarcating supercritical and subcritical flow regimes on the $\mathrm{Pr} -\mathrm{Ta}$ plane. The thick solid and dashed black curves are obtained from the model and DNS respectively. }
\label{FS_two_param}
\end{figure}

Interestingly, even after the drastic simplification, the critical Rayleigh number for the onset of convection determined from the above 1D model is same as the ones determined from the $5$ mode model. 
We have also checked that the one dimensional model (\ref{1deq_f}) provide qualitatively same bifurcation structure as the one given by the $5$ mode model (\ref{5m_f1}) - (\ref{5m_f5}) in the stationary cellular convection regime. Therefore, we use the model (\ref{1deq_f}) for the investigation of subcritical convection and the effect of the parameters on it. Figure~\ref{FS_two_param} shows the regions of subcritical and supercritical onset of convection regimes on the $\mathrm{Pr} - \mathrm{Ta}$ plane obtained from the 1D model separated by the thick solid black curve. The region in the left of the solid curve is for subcritical convection, while, the region on the right side of the curve is for supercritical convection.   

For the validation, next we perform direct numerical simulations of the system in the presence of free-slip boundary conditions using an open source pseudospectral code $Tarang$~\cite{mkv:code}. The simulations are performed in a domain of dimensions $2\pi/k_c \times 2\pi/k_c \times 1$ with $32^3$ spatial grids ($k_c$ is the critical wave number determined from the linear theory). In the code, the independent convective fields are expanded using the Fourier basis functions as
\begin{widetext}
\begin{eqnarray}
(u_z, \theta)&=&\sum_{l,m,n}\left(W_{lmn}(t), T_{lmn}(t)\right) e^{i(lk_xx+mk_yy)}\sin{(n\pi z)},\\
(u_x, u_y)&=&\sum_{l,m,n}\left(U_{lmn}(t), V_{lmn}(t)\right) e^{i(lk_xx+mk_yy)}\cos{(n\pi z)},
\end{eqnarray}
\end{widetext}
where $U_{lmn}$, $V_{lmn}$, $W_{lmn}$ and $T_{lmn}$ are the Fourier modes amplitudes with $l$, $m$, and $n$ being the non-negative integers. $k_x$ and $k_y$ are the horizontal wave numbers along $x$ and $y$ directions respectively such that $k_c^2=k_x^2+k_y^2$. Time advancement is done using the fourth order Rounge-Kutta (RK4) scheme with CFL condition considering maximum time step $\Delta t=0.001$. 

Using the above procedure, we perform extensive DNS of the system and determine the boundary delimiting the regions of supercritical and subcritical onset of convection on the $\mathrm{Pr} - \mathrm{Ta}$ plane. Note that the subcritical and supercritical onset of convection are determined by computing the Nusselt number with the forward and backward variation of the Rayleigh number around the critical Rayleigh number for the onset of convection as was done for the rigid boundary conditions. The boundary is shown with a dashed black line in the figure~\ref{FS_two_param}. The difference between the boundaries delimiting the subcritical and supercritical regions obtained from the model and DNS is more here compared to the rigid boundary case. The reason may be attributed to the drastic simplification of the system. Nonetheless, the flow patterns observed at the supercritical and subcritical regimes are similar both in model and DNS.  It is clearly understood from the model that like rigid boundary conditions,the finite amplitude solutions observed at the onset of convection is of subcritical origin. Much like the rigid boundary conditions, the rotation rate in this case also appears to promote the subcritical convection, while, Prandtl number inhibits it. Thus, the change of boundary conditions does not bring qualitative change on the onset of convection.

\subsection{Overstability}
The five mode models (\ref{5eq1}) - (\ref{5eq5}) and (\ref{5m_f1}) - (\ref{5m_f5}) , not only help to provide the simplest possible descriptions of the stationary supercritical and subcritical onset of rotating convection in terms of one dimensional models (\ref{1deq}) and (\ref{1deq_f}), but also captures the phenomenon of overstable convection leading to small amplitude time dependent periodic solution near the onset of convection. In this section, we utilize the five mode models to investigate the onset of overstable convection. First we use the critical wave number for the onset of overstable convection ($\mathrm{k}_o$) obtained from linear theory both for rigid and free-slip boundary conditions, and determine the critical Rayleigh number for the onset of overstable convection ($\mathrm{Ra}_o$) in the $5$ dimensional models and DNS. The list of values of $\mathrm{Ra}_o$ obtained from linear theory, $5$ mode models, and DNS are presented in the table~\ref{table:comparison_O} and compared for two Prandtl numbers. Interesting to note here that in the entire overstable regime determined from the linear theory (see figure~\ref{marginal_curve}), the $5$ mode models as well as DNS exhibit periodic solution of overstable origin at the onset of convection. 
 
\begin{table*}
\caption{Critical Rayleigh numbers ($\mathrm{Ra_o}$) and wave number ($\mathrm{k_o}$) at the onset of overstability computed from the linear theory (LT), 1D model and DNS for rigid and free-slip boundary conditions.}\label{table:comparison_O}
\begin{tabularx}{\textwidth} { 
   >{\centering\arraybackslash}X 
   >{\centering\arraybackslash}X 
   >{\centering\arraybackslash}X
   >{\centering\arraybackslash}X
   >{\centering\arraybackslash}X
   >{\centering\arraybackslash}X
   >{\centering\arraybackslash}X
   >{\centering\arraybackslash}X
   >{\centering\arraybackslash}X }
 \hline
 \hline
$\mathrm{Pr}$ & $\mathrm{Ta}$ & $\mathrm{k_o}$& $\mathrm{Ra_o}$  & $\mathrm{Ra_o}$ & $\mathrm{Ra}_c$ & Error(\%) & Error(\%) \\
            &   & (LT)  & (LT) & (Model) & (DNS) & (LT vs Model) & (LT vs DNS)\\
\hline
0.1&$2\times 10^4$ & $3.513$ & $6418$    &  $5803$  & $6360$ & $9.59$  & 0.90\\    
(Rigid)&$3\times 10^4$ & $3.662$ & $6978$    &  $6142$  & $7000$ & $11.98$ & 0.31  \\ 
&$5\times 10^4$ & $3.901$ & $7938$    &  $6735$  & $7955$ & $15.15$ &  0.21      \\ 
\hline 
0.025&$1\times 10^4$ & $2.983$ & $4365$    &  $4252$  & $4370$ & $3.14$   & 0.11\\     
(Rigid)&$2\times 10^4$ & $2.996$ & $4507$    &  $4286$  & $4518$ & $4.90$  & 0.24\\ 
&$5\times 10^4$ & $3.036$ & $4812$    &  $4396$  & $4860$ & $8.64$  &  0.99\\ 
\hline
\hline
0.1&$1\times 10^3$ & $2.261$ & $1482$    &  $1484$  & $1485$ & $0.13$ & 0.20\\ 
(Free-slip)&$5\times 10^3$ & $2.401$ & $1614$    &  $1615$  & $1615$ & $0.06$ & 0.06\\
&$1\times 10^4$ & $2.543$ & $1760$    &  $1762$  & $1765$ & $0.11$ & 0.28\\
\hline
0.025&$1\times 10^3$ & $2.224$ & $1350$    &  $1351$  & $1352$ & $0.07$ & 0.14 \\
(Free-slip)&$5\times 10^3$ & $2.237$ & $1360$    &  $1361$  & $1362$ & $0.07$ & 0.14\\
&$1\times 10^4$ & $2.250$ & $1372$    &  $1373$  & $1375$ & $0.07$ & 0.21\\
\hline
\hline
\end{tabularx}
\end{table*}

\begin{figure}
\centering
\includegraphics[scale = 0.55]{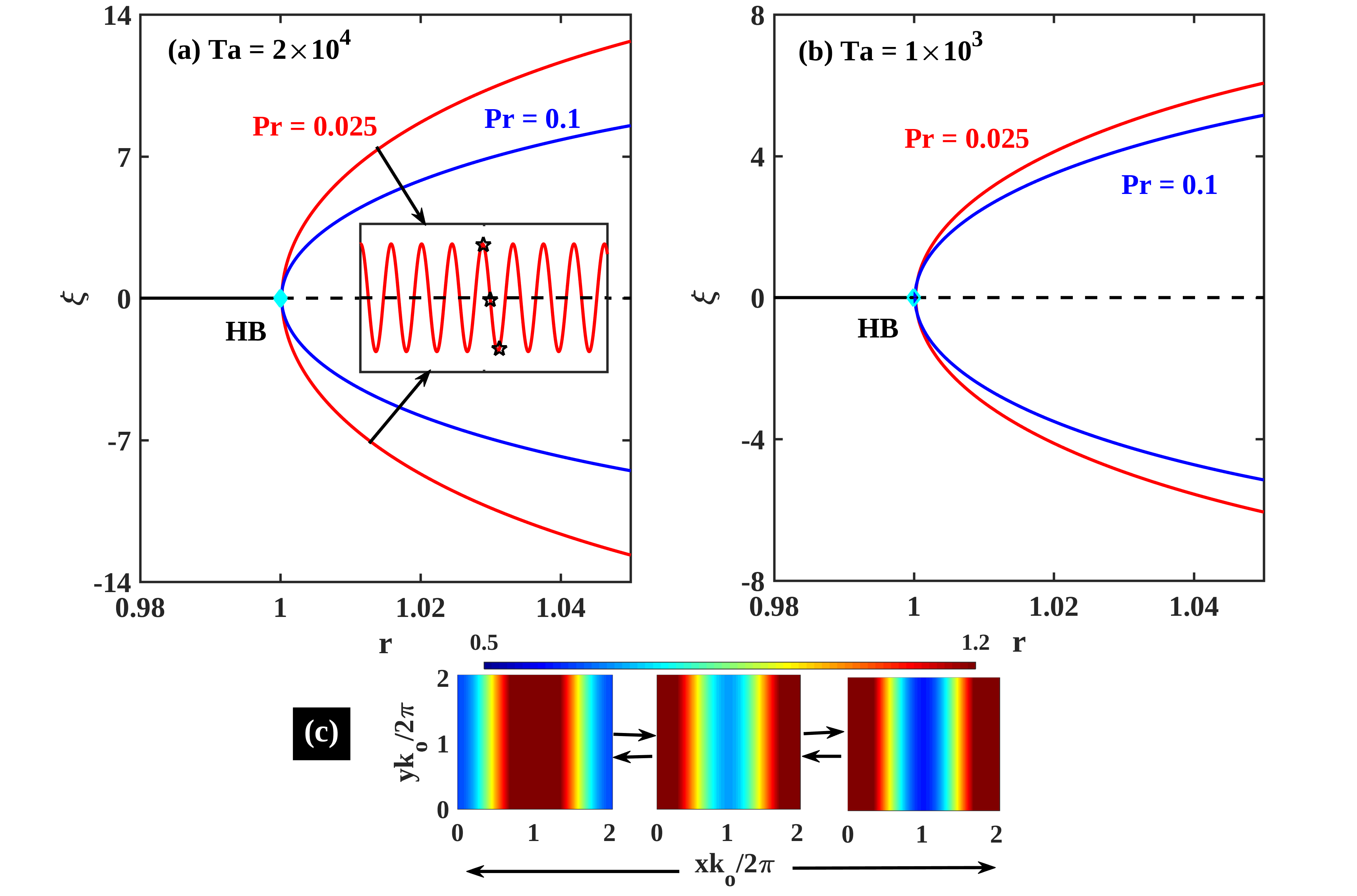}
\caption{Bifurcation diagrams constructed from the 5D models for two $\mathrm{Pr}$ values by plotting the extremum values of the variable $\xi$ corresponding to different solutions with the variation of $\mathrm{r}$. Stable and unstable solutions are represented by solid and dashed lines, respectively. The cyan filled diamond at $r=1$ indicates the supercritical Hopf bifurcation (HB) point. The solid black curve represents the conduction state, while, the red and blue curves correspond to the stable limit cycles.  In (a) and (b) rigid and free-slip boundary conditions are considered respectively.  The inset in (a) displays the time evolution of $\xi$ corresponding to the limit cycle solutions for $\mathrm{Pr}=0.025$. (c) Isotherms computed at the mid plane corresponding to the marked points in the inset of (a).}
\label{overstability_bif}
\end{figure}

Inspired by the above observation, we move ahead and construct bifurcation diagrams using both the $5$ mode models for two different Prandtl numbers ($\mathrm{Pr} = 0.025,~0.1$). We choose $\mathrm{Ta} = 2\times 10^4$ for rigid boundary conditions, while, for free-slip boundary conditions we take $\mathrm{Ta} = 10^3$. The parameters are chosen in such a way that the oscillatory mode of convection of overstable origin is observed at the onset. The bifurcation diagrams presented in the figures~\ref{overstability_bif}(a) and (b) clearly show the birth of a small amplitude oscillatory through supercritical Hopf bifurcation. The temporal evolution of the flow patterns corresponding to oscillatory solutions for rigid boundary conditions are shown in the figure~\ref{overstability_bif}(c). The flow patterns corresponding to the oscillatory solutions of overstable origin for free-slip boundary conditions are also similar. We have checked that the  flow patterns obtained from the DNS, both for rigid and free-slip boundary conditions are also similar. Thus, the $5$ dimensional models qualitatively captures the overstable convection in the considered parameter regime.  

\section{Conclusions}
In summary, we have investigated the transition to convection in rotating Rayleigh-B\'enard convection with rigid and free-slip boundary conditions by performing low dimensional modeling and direct numerical simulations in the Prandtl number range $0 < \mathrm{Pr} \leq 0.6$. The Taylor number is varied in the ranges $0 < \mathrm{Ta} \leq 5\times 10^4$ and $0 < \mathrm{Ta} \leq 10^4$ for rigid and free-slip boundary conditions respectively. 

Extensive three dimensional direct numerical simulations performed  with both rigid and free-slip boundary conditions in the considered parameter regime reveals stationary as well as oscillatory flow patterns at the onset of convection which can be of subcritical and supercritical origin. The supercritical flow regime is characterized by the appearance of small amplitude two dimensional rolls solutions at the onset. On the other hand, finite amplitude stationary two dimensional flow patterns are manifested at the onset of  subcritical convection. For the time dependent flow regime of overstable origin, small amplitude oscillatory solutions are observed at the onset.   

For the simplified mathematical description of the above observations, we perform low dimensional modeling of the system. To our surprise, this lead to the simplest possible description of the stationary supercritical and subcritical rotating convection in terms of an one dimensional model both for rigid and free-slip boundary conditions. The bifurcation analysis of the one dimensional models show that the supercritical and subcritical flow regimes are associated with the pitchfork bifurcations of similar type. The models are then used to identify different stationary flow regimes on the $\mathrm{Pr} - \mathrm{Ta}$ plane and compared with the ones obtained from the DNS. Both the 1D models  for rigid and free-slip boundary conditions show qualitative match with the DNS results, in spite of drastic simplification. Interestingly, a better match is observed with the DNS results for the model with rigid boundary conditions. Moreover, irrespective of the boundary conditions, the finite amplitude solution associated with the subcritical pitchfork bifurcation, is dominantly observed for low Prandtl number fluids in the considered range of the Taylor number in this paper. The Taylor number is found to promote subcriticality, while, the Prandtl number inhibits it. This observation is consistent with the results of the previous numerical simulations~\cite{clever1:JFM_1979,mandal:POF_2022}. 

Further, we also derive two $5$ dimensional models to study the overstable onset of convection in the presence of rigid and free-slip boundaries. The models nicely explains the origin of small amplitude time dependent flow patterns in the region of overstable convection determined from the linear theory. 
The results presented in the paper show that in spite of very high complexity of the RRBC system, the effective dynamics of the system close to the onset can be captured by a very simple set of ordinary differential equations. We expect that similar analysis will be helpful to gain insight about the dynamics of different complex systems including thermal convection in simultaneous presence of rotation and external magnetic field.  

\begin{acknowledgments}
SM and SS acknowledge the supports from CSIR India (File No. 09/973(0024)/2019-EMR-I) and UGC India (Award No. 191620126754) respectively. The authors thankfully acknowledge the suggestions of Manojit Ghosh in constructing the low dimensional models.  
\end{acknowledgments}

\end{document}